\documentclass[structabstract]{aa}  
\usepackage{graphicx}
\usepackage{subfig}
\usepackage{natbib}
\usepackage{txfonts}
\newcommand{\caII}{Ca\,{\scriptsize II}}

\newcommand{\mgII}{Mg\,{\scriptsize II}}
\newcommand{\cII}{C\,{\scriptsize II}}
\def\Mo{M_{\odot}}
\begin{document}

\title{Chromosphere of K giant stars}
\subtitle{Geometrical extent and spatial structure detection}

\author{P.~Berio\inst{1} \and 
			T.~Merle\inst{2} \and 
			F.~Th\'evenin\inst{2} \and
			D.~Bonneau\inst{1} \and
			D.~Mourard\inst{1}  \and
			O.~Chesneau\inst{1} \and
			O.~Delaa\inst{1} \and
			R.~Ligi\inst{1} \and
			N.~Nardetto\inst{1} \and
			K.~Perraut\inst{3} \and
			B.~Pichon\inst{2} \and
			P.~Stee\inst{1} \and
			I.~Tallon-Bosc\inst{4} \and
			J.M.~Clausse\inst{1} \and
			A.~Spang\inst{1} \and
			H.~McAlister\inst{5}\inst{6} \and
			T.~ten Brummelaar\inst{5} \and
			J.~Sturmann\inst{5} \and
			L.~Sturmann\inst{5} \and
			N.~Turner\inst{5} \and
			C.~Farrington\inst{5} \and
			P.J.~Goldfinger\inst{5}
			}
	\institute{
		Nice Sophia-Antipolis University, CNRS UMR 6525, Observatoire de la C\^{o}te dAzur, BP 4229, 06304 Nice Cedex 4, 
France\label{inst1}
		\and Nice Sophia-Antipolis University, CNRS UMR 6202, Observatoire de la C\^{o}te dAzur, BP 4229, 06304 Nice Cedex 4, 
France\label{inst2}
	\and Universit\'e Joseph-Fourier, Institut de Plan\'etologie et d Astrophysique (IPAG) UMR 5274 CNRS, BP 53, 38041 Grenoble Cedex
09, France \label{inst4}
	\and Universit\'e de Lyon, Lyon, F-69003, France ; Universit\'e Lyon 1, Observatoire de Lyon, 9 avenue Charles Andr\'e, Saint-Genis Laval, F-69230, France ; CNRS, UMR 5574, Centre de Recherche Astrophysique de Lyon ; Ecole Normale Sup\'erieure de Lyon, Lyon, F-69007, France\label{inst3}
	\and CHARA Array, Mount Wilson Observatory, 91023 Mount Wilson CA, USA\label{inst5}
	\and Georgia State University, P.O. Box 3969, Atlanta GA 30302-3969, USA\label{inst6}
}

\date{accepted August 31, 2011}

\abstract{Interferometers provide accurate diameter measurements of stars by analyzing both the continuum and the lines formed in photospheres and chromospheres. Tests of the geometrical extent of the chromospheres are therefore possible by comparing the estimated radius in the continuum of the photosphere and the estimated radii in chromospheric lines.}
{We aim to constrain the geometrical extent of the chromosphere of non-binary K giant stars and detect any spatial structures in the chromosphere.}
{We performed observations with the CHARA interferometer and the VEGA beam combiner at optical wavelengths. We observed seven non-binary K giant stars ($\beta$ and $\eta$ Cet, $\delta$ Crt, $\rho$ Boo, $\beta$ Oph, $109$ Her and $\iota$ Cep). We measured the ratio of the radii of the photosphere to the
chromosphere using the interferometric measurements in the H$_{\alpha}$ and the \caII\ infrared triplet line cores. For $\beta$~Cet, spectro-interferometric observations are compared to an non-local thermal equilibrium (NLTE) semi-empirical model atmosphere including a chromosphere. The NLTE computations provide line intensities and contribution functions that indicate the relative locations where the line cores are formed and can constrain the size of the limb-darkened disk of the stars with chromospheres. We measured the angular diameter of seven K giant stars and deduced their fundamental parameters: effective temperatures, radii, luminosities, and masses. We determined the geometrical extent of the chromosphere for four giant stars ($\beta$ and $\eta$ Cet, $\delta$ Crt and $\rho$ Boo). }
{The chromosphere extents obtained range between $16\%$ to $47\%$ of the stellar radius. The NLTE computations confirm that the {\caII}/849 nm line core is deeper in the chromosphere  of $\beta$~Cet than either of the {\caII}/854 nm and {\caII}/866 nm line cores. We present a modified version of a semi-empirical model atmosphere derived by fitting the \caII\ triplet line cores of this star. In four of our targets, we also detect the signature of a differential signal showing the presence of asymmetries in the chromospheres.}
{It is the first time that geometrical extents and structure in the chromospheres of non-binary K giant stars are determined by interferometry. These observations provide strong constrains on stellar atmosphere models.}


\maketitle
%

\section{Introduction}
\begin{table*}[t!]
\caption{List of the K giant stars observed with CHARA/VEGA. The magnitudes in the K band come from the 2MASS catalog \citep{2mass06}. Atmospheric parameters ($T_{\rm eff}$, 
$\log g$ and [Fe/H]) come from \cite{luck95} and \cite{Will90}. $A_V$ is the interstellar absorption in the V band. $p$ is the corrected Hipparcos parallax \citep{vanleeuwen}.} 
\label{table:1} 
\centering 
\begin{tabular}{c c c c c r c l c r c} 
\hline\hline 
HD & Name & Spectral Type & $p$ &$m_V$ & $m_K$ & $V-R$ & $T_{\rm eff}$ & $\log g$ & [Fe/H] & A$_{\rm v}$ \\ 
  &   &   & $(mas)$ &  &   &   & $(K)$ &   &   &   \\ 
\hline 
4128   & $\beta$ Cet  & K0III & $33.86\pm 0.16$ & 2.02 &$-0.20$& 0.72 & $4750\pm 100$ & 2.45 &   0.13 & 0\\ 
6805   & $\eta$ Cet   & K1III & $26.32\pm 0.14$ & 3.45 &  0.86 & 0.83 & $4425\pm 100$ & 2.05 &   0.04 & 0\\
98430  & $\delta$ Crt & K0III & $17.56\pm 0.19$ & 3.56 &  0.95 & 0.83 & $4500\pm 120$ & 2.59 & $-0.48$& 0.01\\
127665 & $\rho$ Boo   & K3III & $20.37\pm 0.18$ & 3.59 &  0.59 & 0.92 & $4260\pm 120$ & 2.22 & $-0.17$& 0\\
161096 & $\beta$ Oph  & K2III & $39.85\pm 0.17$ & 2.77 &  0.27 & 0.82 & $4475\pm 100$ & 1.70 &   0.00 & 0.05\\
169414 & 109 Her      & K2III & $27.42\pm 0.40$ & 3.84 &  1.09 & 0.85 & $4450\pm 120$ & 2.67 & $-0.16$& 0.04\\
216228 & $\iota$ Cep  & K0III & $28.29\pm 0.10$ & 3.53 &  1.11 & 0.83 & $4770\pm 120$ & 2.97 & $-0.12$& 0\\
\hline 
\end{tabular}
\end{table*}

Chromospheres are the outer region of stars characterized by a positive temperature gradient and a departure from radiative equilibrium. 
These properties are caused by a heating mechanism in the low density part of stellar atmospheres, which results in temperatures ranging between the minimum of the temperature profile and $\sim 20,000$~K. 
The sources of this heating have not been tightly constrained but surface convection and magnetic fields seem to be the main mechanisms responsible for heating the upper atmospheres of giant stars hotter than K2. 
For cooler giants ($V-R > 0.8$), these chromospheres look more extended relative to the stellar radius than for hot giants ($V-R < 0.8$) and might be associated with stellar winds and pulsation mechanisms, e.g. mass loss driven by accoustic waves. 
These two groups of giant stars were proposed for the first time by \cite{linsky79} from an analysis of ultraviolet spectra. 
The group of coronal stars (giant stars hotter than K2) display emission lines that must origninate in chromospheres, transition regions, and by implication coronae. The group of non-coronal stars (giant stars cooler than K2) display emission lines formed at temperatures cooler than $20,000$~K, which can originate only in chromospheres.
The Linsky-Haisch dividing line separates the groups of coronal and non-coronal stars in the Herzsprung-Russell (HR) diagram and was extensively studied in the 80's \citep{simon82,haisch87, brown84,carpenter85}. Even if this division of giant stars into two groups seemed too simplistic, hybrid giant stars with mixed coronae and significant wind activity have been revealed (\citealt{reimers82}, \citealt{hall2008}, and \citealt{ayres2010}). For example, the K3III giant star, $\delta$~And, shows an unexpected presence of C IV in emission, which implies that it contains hot material (about 100,000 K), and evidence of a strong, high-velocity wind \citep{judge1987}. This hybrid state appears to represent the transition between coronal and non-coronal groups.\\
Whatever the classification into either coronal, non-coronal, or hybrid groups, no real agreement has been reached about the geometrical extent of the chromosphere of cool giant stars. 
Finally, \cite{judge87} argued the chromosphere extents do not change dramatically as a star crosses the Linsky-Haisch dividing line. 
This result is also supported by model atmospheres developed by \citet{cuntz90A,cuntz90B}, which predict their extent to be between 5$\%$ and 50$\%$ of the stellar radius 
(see Table~3 of \citealt{cuntz90A}). 
These model atmospheres are constructed using spectroscopic observations of \cII, \mgII, and \caII\ lines in the UV and IR wavelength ranges, forcing the radiative transfer of these lines to fit their cores formed in the chromosphere, transition, or coronae regions.\\
It is apparent that the extents of the chromosphere and its physical conditions are still not fully understood for cool giant stars.
The geometrical constraints on observed chromospheres are non-existant except for eclipsing binaries for which the radius can be measured in a spectral line as for example for the $\zeta$ Aur system \citep{eaton93}. 
It is therefore worthwhile to do interferometric observations of red giant stars in the cores of lines formed in chromospheres and to compare the resulting measured radii with those obtained from the continuum during the same run of observations. 
For this purpose we need an interferometer working in the visible or near-IR such as CHARA/VEGA \citep{mourard09} allowing us to observe the H$_{{\rm \alpha}}$ line, the \caII\ triplet lines (849, 855 and 860~nm), and the nearby continuum.

This work compares the measured radii in different spectral lines and the continuum for seven K giant stars, one of these stars being classified as a coronal star ($\beta$~Cet).
The observations and data processing are described in Sect.~2. In Sect.~3, we present our estimates of the fundamental parameters of the program stars, followed in Sect.~4 by our study of the chromosphere of four giant stars. 
In Sect.~5, a semi-empirical model of the chromosphere of $\beta$~Cet is used to interpret spectroscopic and interferometric data. 
Conclusions are presented in the last section.

\section{Interferometric observations}
\subsection{Data}
\label{secdata}
Seven K giant stars, listed in Table~\ref{table:1}, were observed at medium and high spectral resolution with the Visible spEctroGraph And polarimeter (VEGA, \citealt{mourard09}) integrated within the CHARA array at Mount Wilson Observatory (California, USA, \citealt{brummelaar05}). 
We used two criteria to select the giant stars for the program: 
\begin{itemize}
\item the visibility in the stellar continuum at $790$~nm had to be greater than 30\%; 
\item the stars had to be bright enough, i.e. $m_V<4$ (in high spectral resolution mode). 
\end{itemize} 
These two criteria ensured the optimal operation of VEGA when no fringe tracker is used.

Apart from $\beta$ Ophicus,  $\iota$ Cephei, and 109 Herculis, which were observed in the continuum only, all stars were observed in both the continuum and 
chromospheric spectral lines. We selected four lines with chromospheric cores: the H$_{\alpha}$ Balmer line and the \caII\ infrared tripet lines ($849.8$~nm, $854.2$~nm and $866.2$~nm). 
Details of the observations can be found in Table~\ref{table:2}.\\
The shortest baseline of the CHARA array (S1S2, $33$~m) was used during the observations and the data were recorded in two spectral bands of the continuum simultaneously using the two detectors of VEGA (blue and red detectors). 
Calibrator stars were also observed in the continuum in order to calibrate the measurements. 
We used the following sequence of observations {\it cal-target-cal-target-cal} when only one calibrator was available and the sequence {\it cal1-cal2-target-cal2-cal1-target-cal1-cal2} when two calibrators were available. 
We used the medium spectral resolution mode of VEGA ($R=5,000$) for the observations in the continuum.

\begin{table}
\caption{Journal of the observations. $\lambda_0$ is the central wavelength of the recorded band. The last column presents the spectral bandwidth $\Delta\lambda$ used in the data processing (see Sect.~\ref{seqdataproc}).}
\label{table:2} 
\centering 
\begin{tabular}{c c c r c} 
\hline\hline 
Name         & Type   & Epoch      &  $\lambda_0$      & $\Delta\lambda$\\ 
           &        &            &  (nm)   & \\ \hline 
$\beta$ Ceti & target & $10/08/26$ &  $790.0$   & $10$ nm\\ 
           &        & $10/08/26$ &  $620.0$   & $15$ nm\\ 
           &        & $10/07/31$ & \caII\ $849.8$ & $0.25\ \AA$\\ 
           &        & $10/07/31$ & \caII\ $854.2$ & $0.25\ \AA$\\ 
           &        & $10/07/31$ & \caII\ $866.2$ & $0.25\ \AA$\\ 
HD225132     & cal.   & $10/08/26$ &  $790.0$   & $10$ nm\\ 
           &        & $10/08/26$ &  $620.0$   & $15$ nm\\ 
\hline
$\eta$ Ceti  & target & $10/09/16$ &  $790.0$   & $10$ nm\\
           &        & $10/09/16$ &  $620.0$   & $15$ nm\\
           &        & $10/09/17$ & \caII\ $849.8$ & $1\AA$\\ 
           &        & $10/09/16$ & \caII\ $854.2$ & $1\AA$\\ 
           &        & $10/09/17$ & \caII\ $866.2$ & $1\AA$\\ 
HD222345     & cal.   & $10/09/16$ &  $790.0$   & $10$ nm\\ 
           &        & $10/09/16$ &  $620.0$   & $15$ nm\\ 
\hline
$\delta$ Crateris & target & $10/05/05$ &   $790.0$  & $10$ nm\\
                &        & $10/05/05$ &   $620.0$  & $15$ nm\\
                &        & $10/05/05$ & \caII\ $854.2$& $1\AA$\\
                &        & $10/05/05$ & H$_\alpha$ $656.2$ & $1\AA$\\
HD100889          & cal.   & $10/05/05$ &   $790.0$  & $10$ nm\\ 
                &        & $10/05/05$ &   $620.0$  & $15$ nm\\ 
\hline
$\rho$ Bootis     & target & $10/06/26$ &   $790.0$  & $10$ nm\\
                &        & $10/06/26$ &   $620.0$  & $15$ nm\\
                &        & $10/07/31$ & \caII\ $854.2$ & $1\ \AA$\\
HD143894          & cal.   & $10/06/26$ &   $790.0$  & $10$ nm\\ 
                &        & $10/06/26$ &   $620.0$  & $15$ nm\\ 
HD108382          & cal.   & $10/06/26$ &   $790.0$  & $10$ nm\\ 
                &        & $10/06/26$ &   $620.0$  & $15$ nm\\ 
\hline
$\beta$ Ophicus   & target & $10/09/16$ &   $790.0$  & $10$ nm\\
                &        & $10/09/16$ &   $620.0$  & $15$ nm\\
HD152614          & cal.   & $10/09/16$ &   $790.0$  & $10$ nm\\ 
                &        & $10/09/16$ &   $620.0$  & $15$ nm\\
\hline
109 Herculis      & target & $10/07/31$ &   $790.0$  & $10$ nm\\
                &        & $10/07/31$ &   $620.0$  & $15$ nm\\
HD168151          & cal.   & $10/07/31$ &   $790.0$  & $10$ nm\\ 
                &        & $10/07/31$ &   $620.0$  & $15$ nm\\ 
HD166230          & cal.   & $10/07/31$ &   $790.0$  & $10$ nm\\ 
                &        & $10/07/31$ &   $620.0$  & $15$ nm\\ 
\hline
$\iota$ Cephei    & target & $10/11/10$ &   $790.0$   & $10$ nm\\
                &        & $10/11/10$ &   $620.0$   & $15$ nm\\
HD3360            & cal.   & $10/11/10$ &   $790.0$   & $10$ nm\\ 
                &        & $10/11/10$ &   $620.0$   & $15$ nm\\
\hline 
\end{tabular}
\end{table}

The high spectral resolution mode of VEGA ($R=30,000$) was used for the observations of spectral lines for which calibrators were not required because the visibility in the spectral lines were calibrated by the measurements in the continuum, close to the spectral lines (see Sect.~\ref{seqdataproc}).

We selecte the calibrators using the SearchCal tool\footnote{available at: $http://www.jmmc.fr/searchcal\_page.htm$} developed at JMMC \citep{bonneau06}, providing an estimate of the limb-darkened (LD) angular diameter ($\theta_{LD}$).
The uniform-disk (UD) angular diameter ($\theta_{UD}$) is required to estimate the transfer function of the instrument at each wavelength. 
The UD angular diameter of each calibrator is derived using $\theta_{LD}$ with the linear LD coefficients given by \citet{claret95} and \citet{diaz95}.
For each calibrator, Table~\ref{table:3} presents $\theta_{UD}$ at 620~nm and 790~nm.

\begin{table}
\caption{Limb-darkened $\theta_{LD}$ and uniform-disk $\theta_{UD}$ angular diameters of calibrators.} 
\label{table:3} 
\centering 
\begin{tabular}{c c c c} 
\hline\hline 
Name & $\theta_{LD}$ & $\theta_{UD}$ at 620~nm & $\theta_{UD}$ at 790~nm \\ 
& (mas) & (mas) & (mas) \\ 
\hline 
HD3360   & $0.290\pm 0.020$  & $0.283$  & $0.284$  \\ 
HD166230 & $0.420\pm 0.029$ & $0.393$  & $0.398$  \\ 
HD168151 & $0.679\pm 0.047$ & $0.663$  & $0.670$  \\ 
HD152614 & $0.312\pm 0.022$ & $0.301$  & $0.303$  \\ 
HD108382 & $0.394\pm 0.027$ & $0.376$  & $0.380$  \\ 
HD143894 & $0.379\pm 0.026$ & $0.362$  & $0.365$  \\ 
HD100889 & $0.305\pm 0.021$ & $0.293$  & $0.296$  \\ 
HD222345 & $0.511\pm 0.035$ & $0.485$  & $0.491$  \\ 
HD225132 & $0.351\pm 0.024$ & $0.338$  & $0.341$  \\ 
\hline 
\end{tabular}
\end{table}

\subsection{Data processing}
\label{seqdataproc}
The data processing of VEGA is composed of two parts. First, the data in the stellar continuum (at medium spectral resolution) are processed with  the power spectral method giving the squared visibilities. Second, the processing of the data in spectral lines is based on the cross-spectrum method, which provides differential visibilities and phases across the lines (see \citealt{mourard09} for details).

For this analysis and in the case of observations at medium spectral resolution, we divide the whole spectral band recorded by the red detector into four spectral channels of $10$~nm centered on the wavelengths $775$~nm, $785$~nm, $795$~nm, and $805$~nm. 
For the blue detector, the continuum is visible only at $625.5$~nm, we then use only one spectral channel (with a spectral bandwidth of $15$~nm). 
The processing of these five spectral bands give the squared visibilities used to constrain the LD angular diameter (see Sect. \ref{angdiamsusec}).

\begin{figure}
\vspace{1cm}
\includegraphics[height=60mm]{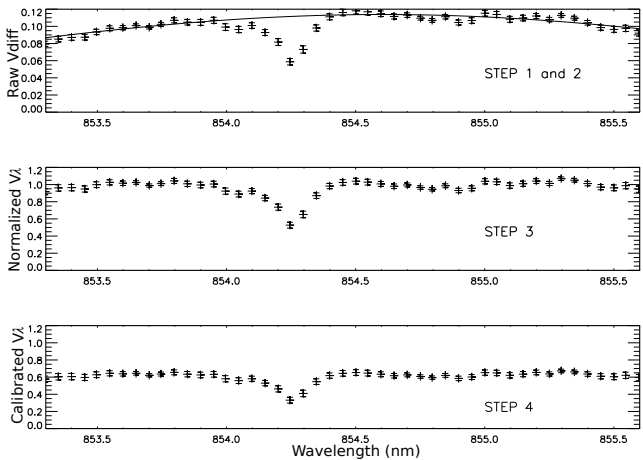}
\caption{Illustration of the four steps of the processing of observations 
at high spectral resolution in an absorption line. 
These data correspond to $\eta$~Cet observed in the \caII/854~nm line. 
Top: raw differential visibility with the simple model. Middle: normalized visibility 
in the sliding narrow spectral band. Bottom: calibrated visibility in the sliding 
narrow spectral band.}
\label{fig1}
\end{figure}

The cross-spectrum method is applied to the data recorded in the high spectral resolution mode. 
We compute the complex differential visibility between a large spectral channel used as reference (centered at $\lambda_1$) and a narrow spectral band (centered at $\lambda_2$) sliding in the reference spectral channel
\begin{equation}
V_{\rm diff}=V_{\lambda_1}V_{\lambda_2}\exp{[-i(\phi_{\lambda_1}-\phi_{\lambda_2})]},
\end{equation}
where $V_{\lambda_i}$ and $\phi_{\lambda_i}$ represent the visibility and the phase of the fringe patterns at the wavelength $\lambda_i$. The width of the sliding narrow spectral band ($\Delta\lambda_2$ hereafter) is chosen to ensure a sufficient signal-to-noise ratio $(S/N)$ for the estimation of the complex differential visibility. 
Therefore, it depends on the stellar magnitude. We use the following spectral bandwidths: $\Delta\lambda_2=0.25\, \AA$ for $\beta$~Cet and $\Delta\lambda_2=1\,\AA$ for $\delta$~Crt, $\rho$~Boo, and $\eta$~Cet.\\
The cross-spectrum method recovers both the modulus of $V_{\rm diff}$ (the differential visibility) and its argument (the differential phase $\Delta\phi_{12}=\phi_{\lambda_1}-\phi_{\lambda_2}$). 
As no calibrator has been observed for the high spectral resolution mode, we developed a specialized processing method to calibrate the visibility in the narrow spectral band. 
The sequence of processing removing the instrumental/atmospheric signature was:
\begin{enumerate}
\item Estimation of $V_{\rm diff}$ using the cross-spectrum method.
\item Fitting of $V_{\rm diff}$ with a simple model as described in \cite{mourard09}. This model describes the instrumental/atmospheric signature.
\item Normalization of $V_{\rm diff}$  with the fitted model. \\

\noindent{We then obtained $V_{\lambda_2}$ normalized to one in the continuum.}\\

\item Using the angular diameter deduced from our observations in the continuum, we computed the theoretical visibility to rescale $V_{\lambda_2}$ (see Fig.~\ref{fig1}). 
\end{enumerate}

\section{Fundamental parameters}
\subsection{Angular diameter}
\label{angdiamsusec}
The squared visibilities in all spectral channels are fitted with a model of LD disk, which is constructed using the laws of \citet{claret95} and \citet{diaz95}. 
Table~\ref{table:4} presents the LD angular diameters $\theta_{LD}$. 
Comparison with a previously published diameter is only possible for $\beta$~Cet for which the diameter was estimated with the instrument VINCI at VLTI in the K band \citep{richichi09}. 
Our estimate for $\beta$~Cet ($5.288\pm0.075$~mas) agrees well with the VINCI one ($5.329\pm 0.005$~mas). 

\begin{table*}
\caption{Fundamental parameters of the observed K giant stars.} 
\label{table:4} 
\centering 
\begin{tabular}{r c c c c c c} 
\hline\hline 
HD &Name &  $\theta_{LD}$ & $f_{\rm bol}$ & $T_{\rm eff}$ & $L/L_\odot$ & $R/R_\odot$ \\ 
 & & (mas) &  ($10^{-6}$~erg~rd$^2$~s$^{-1}$~cm$^{-2}$) & (K) &  & \\ 
\hline 
4128&$\beta$ Ceti      & $5.288\pm 0.075$ & $5.10$ & $4838\pm 70$ & $139.1\pm 7.0$ & $16.78\pm 0.25$ \\ 
6805&$\eta$ Ceti       & $3.698\pm 0.160$ & $1.64$ & $4356\pm 55$ & $~74.0\pm 3.7$ & $15.10\pm 0.10$ \\ 
98430&$\delta$ Crateris & $3.667\pm 0.022$ & $1.66$ & $4408\pm 57$ & $171.4\pm 9.0$ & $22.44\pm 0.28$ \\ 
127665&$\rho$ Bootis     & $4.090\pm 0.031$ & $1.76$ & $4298\pm 56$ & $131.9\pm 6.8$ & $21.57\pm 0.25$ \\ 
161096&$\beta$ Ophicus   & $4.606\pm 0.045$ & $3.04$ & $4621\pm 62$ & $~63.4\pm 3.2$ & $12.42\pm 0.13$ \\ 
169414&109 Herculis      & $3.223\pm 0.034$ & $1.17$ & $4334\pm 59$ & $~50.7\pm 2.7$ & $12.63\pm 0.22$ \\ 
216228&$\iota$ Cephei    & $2.646\pm 0.048$ & $1.28$ & $4831\pm 74$ & $~49.6\pm 2.5$ & $10.05\pm 0.18$ \\ 
\hline 
\end{tabular}
\end{table*}

\begin{figure}
\center{\includegraphics*[height=50mm]{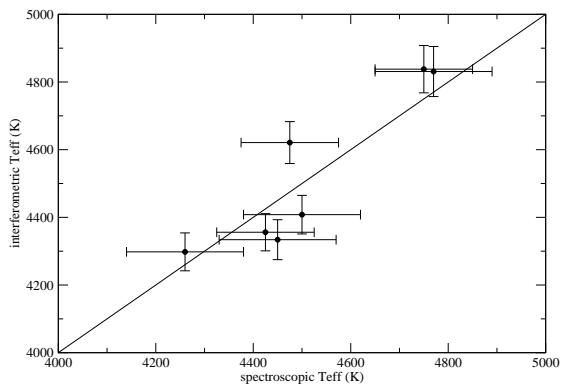}}
\caption{Comparison of spectroscopic and interferometric estimations of $T_{\rm eff}$ for our program stars. The spectroscopic $T_{\rm eff}$ come from Table \ref{table:1}.}
\label{fig2}
\end{figure}

\subsection{Effective temperature}
The effective temperature $T_{\rm{eff}}$ of a star, its LD angular diameter $\theta_{LD}$, and the bolometric flux $f_{{\rm bol}}$ are related via the equation
\begin{equation}
T_{\rm eff}^4=\frac{4f_{{\rm bol}}}{\sigma\theta_{LD}^2},
\end{equation}
where $\sigma$ is the Stefan Boltzmann constant. 
The bolometric fluxes of each star are given in Table~\ref{table:4} and were computed using the bolometric correction in K band from \citet{houdashelt} and the interstellar absorption $A_V$ from the work of \citet{chen98}. 
For all the program stars, it turns out that the absorption is very small because they are not in the direction of the Galactic plane and are closer than 100~pc. 
The interferometric $T_{\rm{eff}}$ obtained are presented in Table~\ref{table:4}. 
We assume an error of $5\%$ in $f_{\rm bol}$, which is associated with the error in the photometry. 
Therefore, the errors in the $T_{\rm{eff}}$ range from $55$~K to $74$~K, implying errors of about $1.5\%$ in the interferometric $T_{\rm{eff}}$.

\begin{figure}[t]
\center{\includegraphics*[height=60mm]{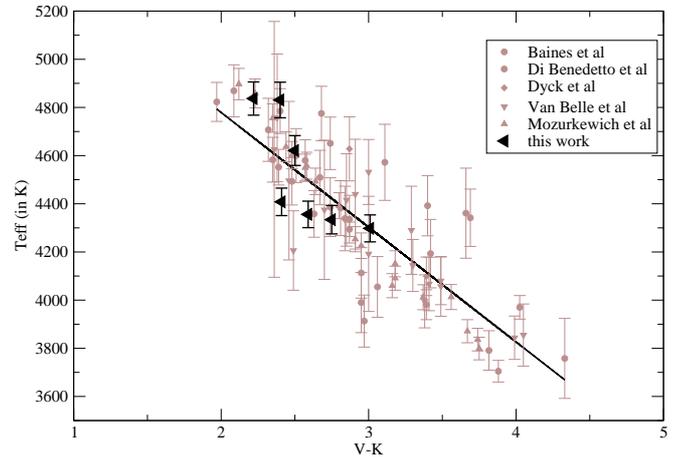}}
\caption{Comparison of our estimates of $T_{\rm eff}$ for our program stars with the temperature scale deduced from other interferometric measurements. }
\label{fig3}
\end{figure}

A comparison of our interferometric $T_{\rm{eff}}$ estimates with the spectroscopic $T_{\rm{eff}}$  found in the literature (\citealt{luck95} and \citealt{Will90}) is presented in Fig.~\ref{fig2} showing that our interferometric measurements are within the spectroscopic error bars.
In Fig.~\ref{fig3}, we show that our $T_{\rm{eff}}$ estimates agree very well with the temperature scale $T_{\rm{eff}}$ versus $(V-K)$ deduced from the interferometric $T_{\rm{eff}}$ estimated by several authors for K giant stars \citep{baines10,mozur03,vanbelle99,dyck96,diben87}. 
This figure shows that our program stars follow the fitted linear-regression relation and that our estimates are within the spread of $\Delta T_{{\rm rms}}\approx 145$~K proposed by \cite{dyck96}.

\subsection{Position in the HR diagram}
\begin{figure}[t]
\center{\includegraphics*[width=85mm]{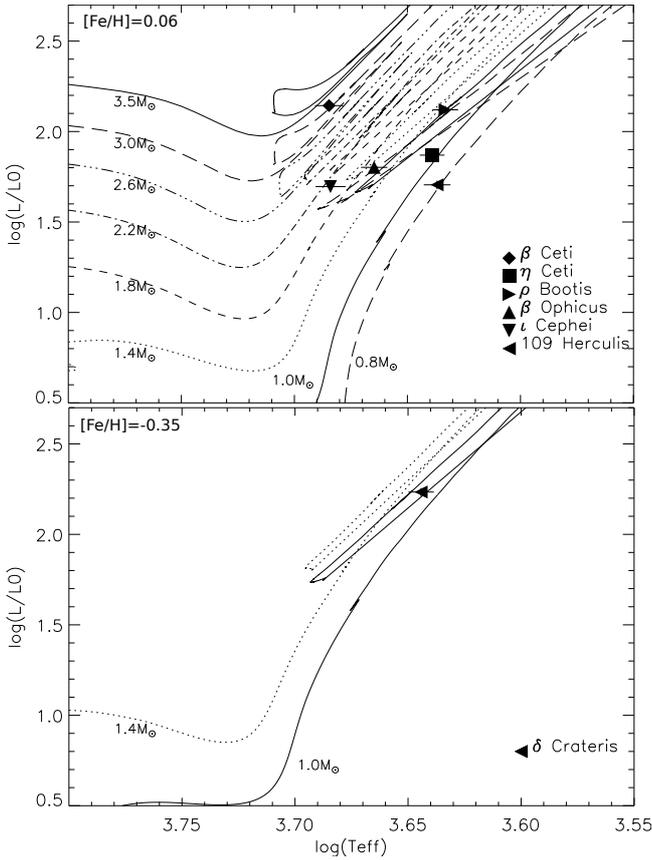}}
\caption{HR diagram of the program stars and comparison with evolutionnary track models for different masses. Top: ${\rm [Fe/H]}=0.06$; bottom: ${\rm [Fe/H]}=-0.35$.}
\label{fig4}
\end{figure}

The radius is obtained with the formula
\begin{equation}
R/R_\odot=\frac{\theta_{LD}}{9.305p},
\end{equation}
where $p$ is the parallax in arcseconds coming from the corrected Hipparcos catalog \citep{vanleeuwen} and $\theta_{LD}$ is in mas. The luminosity is obtained from the absolute bolometric magnitude computed with the apparent magnitude in V band (see Table~\ref{table:1}), the corrected Hipparcos parallax, and the bolometric correction from \citet{houdashelt}.
Results for $L/L_\odot$ and $R/R_\odot$ are given in Table~\ref{table:4}. 
Plotting these results in a HR diagram together with evolutionary tracks  results in a rough estimate of the mass of the program stars (see Fig.~\ref{fig4}). 
We used evolutionary track models at solar metallicity except for $\delta$~Crt for which we use models at [Fe/H]=-0.35, as shown in the bottom panel of Fig.~\ref{fig4}.
The evolutionary tracks come from the BaSTI database \citep{Cassisi04}. 
Table~\ref{TableMass} presents the mass estimates deduced from this comparison. For $\beta$~Cet, we determine a more accurate mass of $3.1M_\odot$ using the CESAM code \citep{morel97}, in good agreement with the results presented in Table~\ref{TableMass}.

\begin{table}
\caption{Estimated masses of the seven observed K giant stars deduced from the BaSTI evolutionnary tracks.} 
\label{TableMass} 
\centering 
\begin{tabular}{c c} 
\hline\hline 
Name &  Mass \\ 
 &  ($M_\odot$)  \\ 
\hline 
$\beta$ Ceti      & 3.0 -- 3.5\\ 
$\eta$ Ceti       & 1.0 -- 1.4\\ 
$\delta$ Crateris & 1.0 -- 1.4\\ 
$\rho$ Bootis     & 1.0 -- 1.4\\ 
$\beta$ Ophicus   & 1.4 -- 1.8\\ 
109 Herculis      & 0.8 -- 1.0\\ 
$\iota$ Cephei    & 1.8 -- 2.2\\ 
\hline 
\end{tabular}
\end{table}

\begin{table*}
\caption{Extent of the chromospheric line-forming regions. The extents are given in UD angular diameter and in stellar radius.} 
\label{tableextension} 
\centering 
\begin{tabular}{c | c c c c | c c c c} 
\hline\hline 
Name &  \multicolumn{4}{c|}{UD Angular diameter ($\theta_{UD}$ in mas)} & \multicolumn{4}{c}{Linear radius (in $R_\star$)} \\
& {\caII}/849 nm & {\caII}/854 nm & {\caII}/866 nm & H$_\alpha$ & {\caII}/849 nm & {\caII}/854 nm & {\caII}/866 nm & H$_\alpha$  \\ 
\hline 
$\beta$ Ceti & $5.82\pm0.20$ & $6.40\pm0.23$ & $6.58\pm0.24$ & & $1.16\pm0.04$ & $1.27\pm0.05$ & $1.31\pm0.05$ & \\
$\eta$ Ceti & $4.40\pm0.19$ & $4.87\pm0.20$ & $5.05\pm0.24$ & & $1.25\pm0.05$ & $1.38\pm0.06$ & $1.43\pm0.07$ & \\
$\delta$ Crateris & & $5.10\pm0.30$ & & $5.13\pm0.18$ & & $1.46\pm0.09$ & & $1.47\pm0.05$ \\
$\rho$ Bootis &  & $4.86\pm0.28$ &  & &  & $1.25\pm0.08$ &  & \\
\hline 
\end{tabular}
\end{table*}

\section{Chromosphere parameters}

\subsection{Geometrical extent}
\begin{figure}[t]
\center{\includegraphics[width=8.5cm]{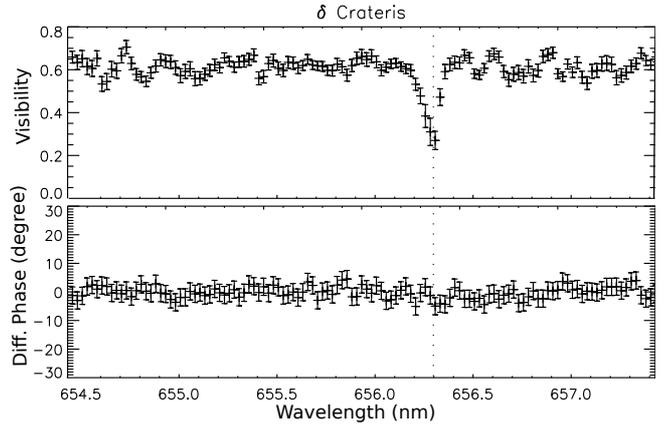}}
\caption{Plots of the visibility and the differential phase versus wavelength in the case of observations around the H$_\alpha$ line (dotted vertical line) for $\delta$~Crateris.}
\label{figdiff656}
\end{figure}

Visibility measurements in the core of the chromospheric lines give us direct estimates of the extent of the line-forming regions. For this purpose, we assume that the line cores are optically thick in the region where they are formed, i.e. the chromosphere. This is demonstrated in Sect.~\ref{secmodel} for $\beta$~Cet for which we have a semi-empirical model of the extended atmosphere. This hypothesis agrees with similar studies for the Sun \citep{white62}. Hence, in a first approximation, we can consider a UD model to derive the extent of the chromosphere.\\
Applying the cross-spectrum method to reduce the data recorded in the high spectral resolution mode (see Sect. \ref{seqdataproc}), we derive the visibility curves as presented in Figs. \ref{figdiff656}, \ref{figdiff854}, \ref{figdiff849}, and \ref{figdiff866}. \\
For each giant star, we then deduce a UD angular diameter from the value of the visibility in the core of the lines. The results are presented in Table~\ref{tableextension}, which also gives the extent in linear stellar radius $R_{\star}$ equal to the ratio of the LD angular diameter in the lines to the stellar LD angular diameter given in Table \ref{table:4}. In a first approximation for computing $\theta_{LD}$ in the core of the lines, we use the same correcting factor $\theta_{LD}/\theta_{UD}\approx 1.05$ given by the limb-darkening laws derived with the semi-empirical model of $\beta$~Cet (see Sect.\ref{secmodel}). \\
In our program stars, only $\beta$~Cet belongs to the group of coronal stars, which means it is located on the left of the Linsky-Haisch dividing line of the HR diagram ($V-R=0.72$). \cite{carpenter85} also give the same classification for $\beta$~Cet. 
We find that the extents of the emitting regions of the \caII\ triplet and H$_\alpha$ lines range from $16\%$ to $47\%$ of the stellar radius. These results show that the extent of the chromospheric line-forming regions is similar for the four cool giant stars regardless of their classification. This conclusion agrees with \cite{judge87}, but not with \cite{carpenter85} who proposed the idea that coronal stars appear to have a very thin chromosphere of no more than $0.1\%$ of the stellar radius. Moreover, our estimated chromosphere sizes  agree well with the theoretical predictions of \cite{cuntz90A}.

\subsection{Structures at the surface}
\begin{figure*}[t]
\center{\includegraphics[width=17cm]{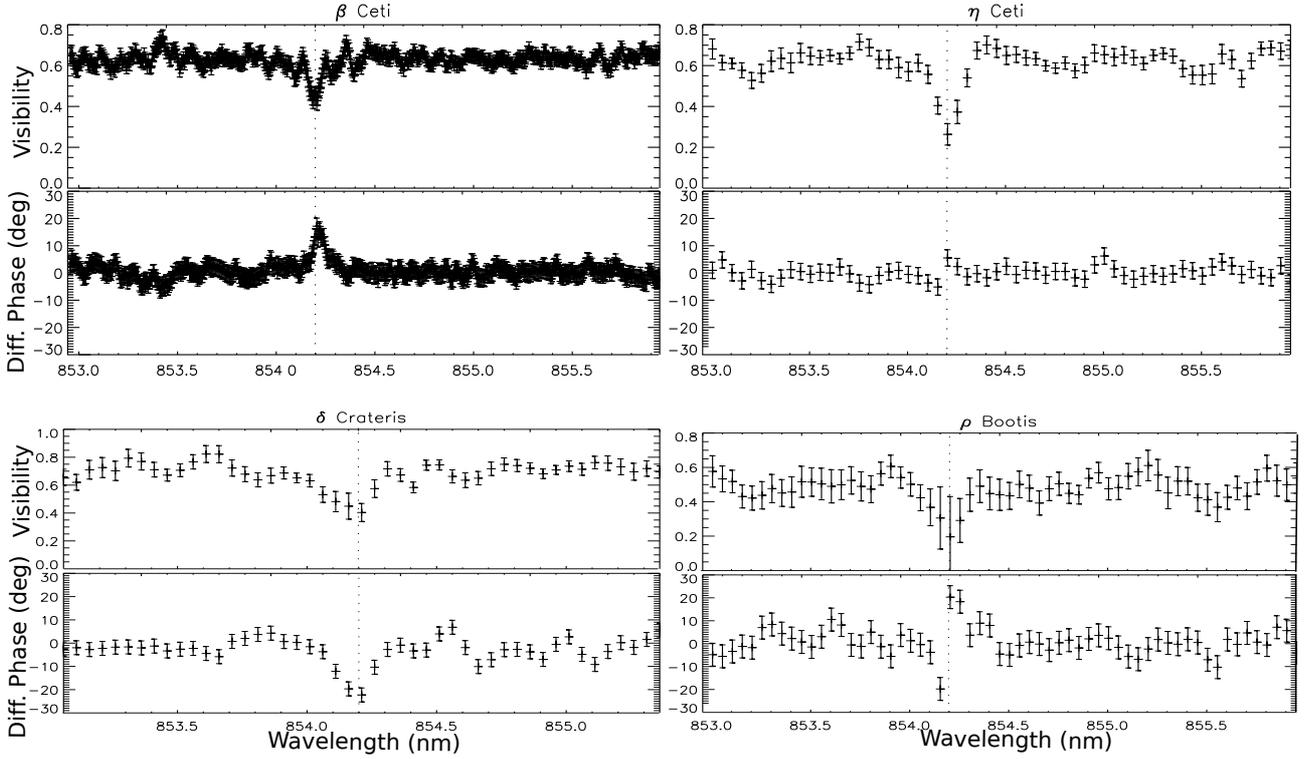}}
\caption{Plots of the visibility and the differential phase versus wavelength in the case of observations around the \caII/854~nm line  (dotted vertical line) for four K giant stars.}
\label{figdiff854}
\end{figure*}

\begin{figure*}[t]
\center{\includegraphics[width=17cm]{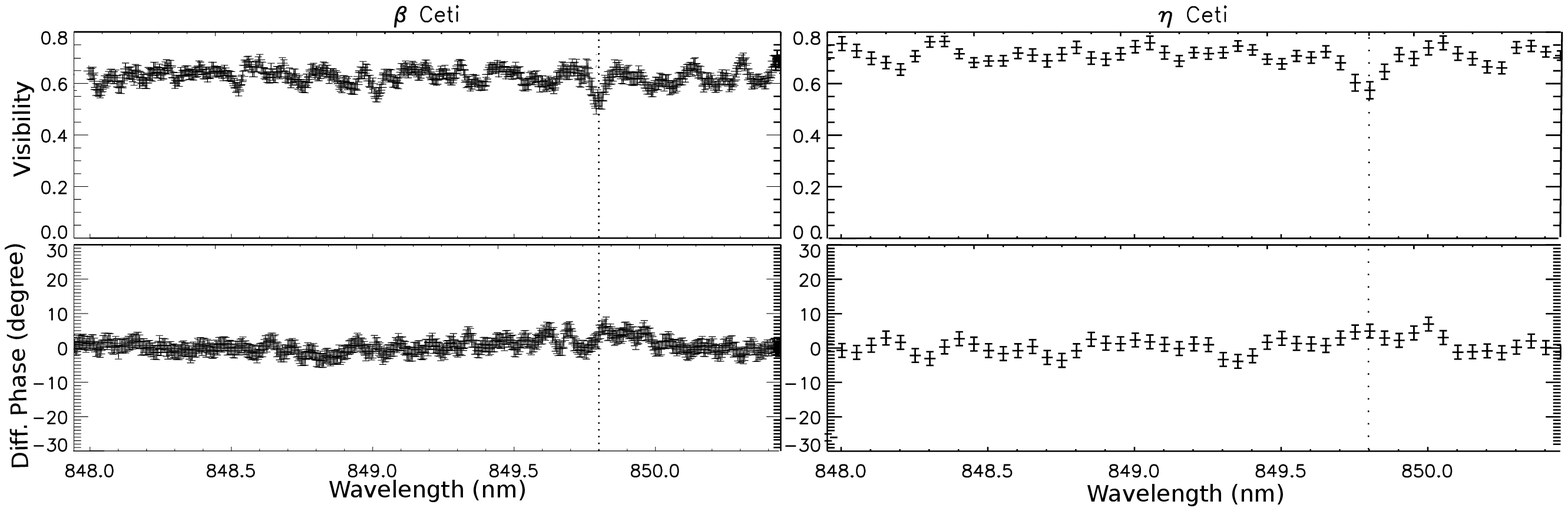}}
\caption{Plots of the visibility and the differential phase versus wavelength in the case of observations around the \caII/849~nm line (dotted vertical line) for two K giant stars.}
\label{figdiff849}
\end{figure*}

\begin{figure*}[t]
\center{\includegraphics[width=17cm]{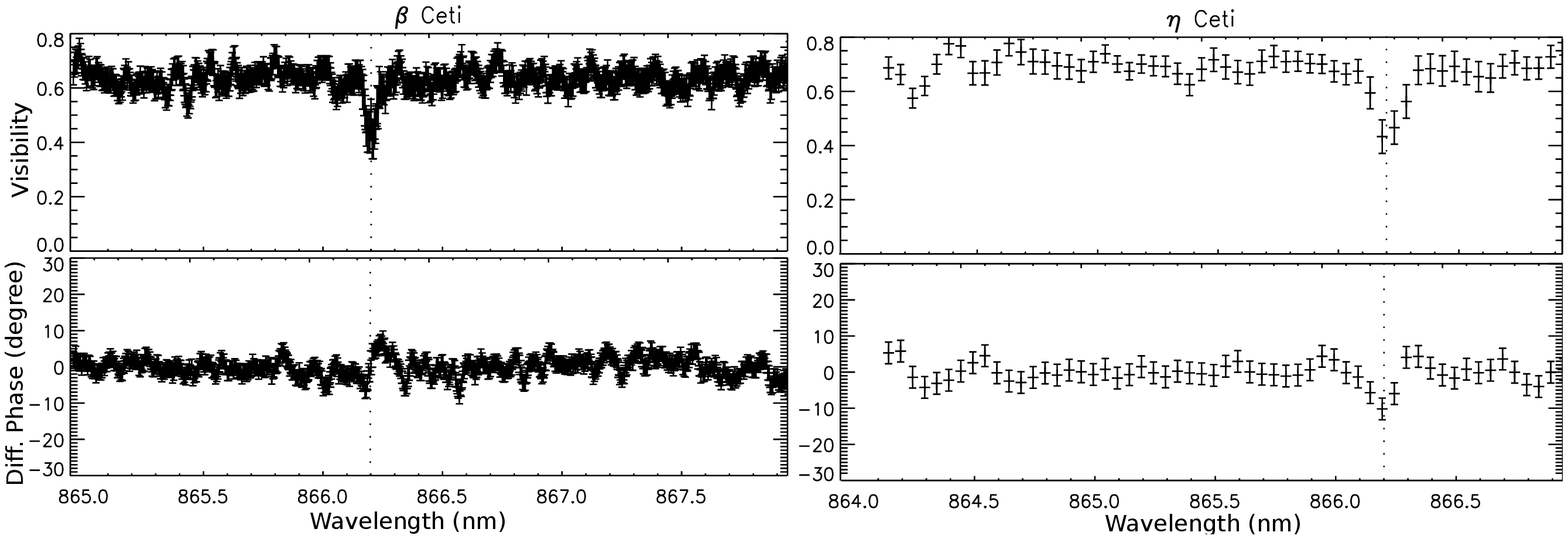}}
\caption{Plots of the visibility and the differential phase versus wavelength in the case of observations around the \caII/866~nm (dotted vertical line) line for two K giant stars.}
\label{figdiff866}
\end{figure*}

Information related to the position of the photocenter of the chromosphere can be deduced by measuring the phase of the visibility (also called differential phase) in the narrow spectral bands. This observable has been intensively used to study the kinematic of circumstellar environments (e.g. \citealt{berio99} and \citealt{meilland07}). In Figs.~\ref{figdiff656}, \ref{figdiff854}, \ref{figdiff849}, and \ref{figdiff866}, we plot the differential phase with respect to the wavelength used for the observation of the chromospheric lines. Clear jumps in the differential phase at the center of the {\caII}/854~nm line are revealed for three of the four program stars observed at high spectral resolution. Only $\eta$~Cet does not exhibit an unambiguous signal. For {\caII}/849~nm and H$_\alpha$ lines, no jump is detected in the differential phase. For the {\caII}/866~nm line, jumps seem to be present but their amplitude with respect to the phase precision is too low for a significant detection. 


The phase jumps can be explained either in terms of an asymmetric chromosphere or an asymmetric photosphere, which implies structure in the chromosphere and/or in the photosphere. 
Our observations provide phase measurements for only one baseline for each line preventing us from deriving the exact photocenter positions. However, the fact that phase signatures are not present with the same amplitude in all lines 
may indicate that structures are present in the chromosphere and not only in the photosphere. The star $\delta$~Crt perfectly illustrates this property because an unambiguous phase jump (amplitude greater than $20^\circ$ seen in the bottom left panel of Fig.~\ref{figdiff854}) is present in the core of the {\caII}/854~nm line whereas the phase remains constant over the H$_{\alpha}$ line (see Fig.~\ref{figdiff656}). 

Until now, only spectroscopic methods have been used to detect and study the presence of structures in the chromosphere of single K giant stars. For instance, \cite{montes00} studied in detail the chromosphere, discriminating between different structures (plages, prominences, flares, and microflares), by analyzing the ratio of excess emission equivalent width (EW) of  two \caII\ infrared lines ($E_{8542}/E_{8498}$) or the ratio of EWs of two Balmer lines ($E_{{\rm H_\alpha}}/E_{{\rm H_\beta}}$). Our study shows that additional informations could be provided by interferometry in the future to help us understand the structures in the chromosphere of single K giant stars. In reality, the structure of the chromospheric network
delineated by bundles of magnetic field
lines could be spatially studied in the future. Combinations of interferometric
observations in several spectral lines will allow us to determine the characteristics (size, position, intensity)
of the chromospheric structures such as plages (part of the chromospheric
network of bright emission associated
with concentrations of magnetic fields) and prominences or
filaments (dense clouds of material suspended above
the stellar surface by loops of magnetic field). These 
perspectives require interferometric observations with
a more complete $(u,v)$ coverage (several baselines with different lengths
and orientations) than that available from these first observations.

\begin{table}
\caption{Comparison of the model atmosphere parameters for $\beta$ Cet. 
        Second column is the model geometry: plane--parallel (pp) or spherical (sph). 
        Last column is the plateau temperature of the chromosphere $T_{\rm p}$.}
\begin{tabular}{cccccc}
\hline \hline 
Model atmosphere &geo.& $T_{\rm eff}$ & $\log g$ & [Fe/H] & $T_{\rm p}$ \\
  & & (K) &   &   & (K) \\\hline
\citet{Eriksson83} &pp & 4900 & 2.90 & 0 & 5500 \\
This work          &sph& 4830 & 2.45 & 0 & 4500 \\
\hline
\end{tabular}
\label{Atm_carac}
\end{table}

\section{Semi-empirical model of $\beta$ Cet}
\label{secmodel}
The star $\beta$~Cet is the only one for which a model atmosphere including a chromosphere and a transition region has been fitted \citep{Eriksson83}. The adopted atmospheric parameters for the target are given in Table~\ref{Atm_carac}. 

We note that \cite{Eriksson83} used a pre-Hipparcos paralax estimate of $p = 61$~mas, which is about twice the Hipparcos measurement ($p = 34$ mas, \citealt{Perryman97}). The star was thought to be closer, thus smaller and consequently leading to an overestimation of the surface gravity by \cite{Eriksson83}. We adopt a surface gravity based on our previously estimated  interferometric radius ($R=16.8\ R_\odot$) and the mass deduced from the evolutionary tracks ($M = 3\ M_\odot$). The uncertainties in these two parameters give a surface gravity ranging from 2.45 to 2.55. We adopted a value of $\log g = 2.45$, which is consistent with the values determined by both \citet{luck95} and \citet{Kovacs83} from a spectroscopic analysis based on the ionization equilibrium of iron. We interpolate the MARCS\footnote{available at: http://marcs.astro.uu.se} model atmospheres \citep{Gustafsson08} with spherical geometry for the atmospherical parameters $T_{\rm eff} = 4830$~K, $\log g =2.45$, [Fe/H] = 0, and $\zeta = 2$~km~s$^{-1}$, all obtained with the interpolation code of T.~Masseron\footnote{available at: http://marcs.astro.uu.se/software.php}. For the interpolation, we note that we use four spherical model atmospheres corresponding to a stellar mass of 2~$\Mo$ available in the MARCS grids.
The theoretical MARCS model atmospheres do not include either chromosphere or transition region. To reproduce the core of the \caII\ triplet lines, we use the model of \citet{Eriksson83} replacing its photospheric part by the interpolated MARCS model, hereafter called  the hybrid model.

\begin{figure}[]
\includegraphics[width=8.5cm]{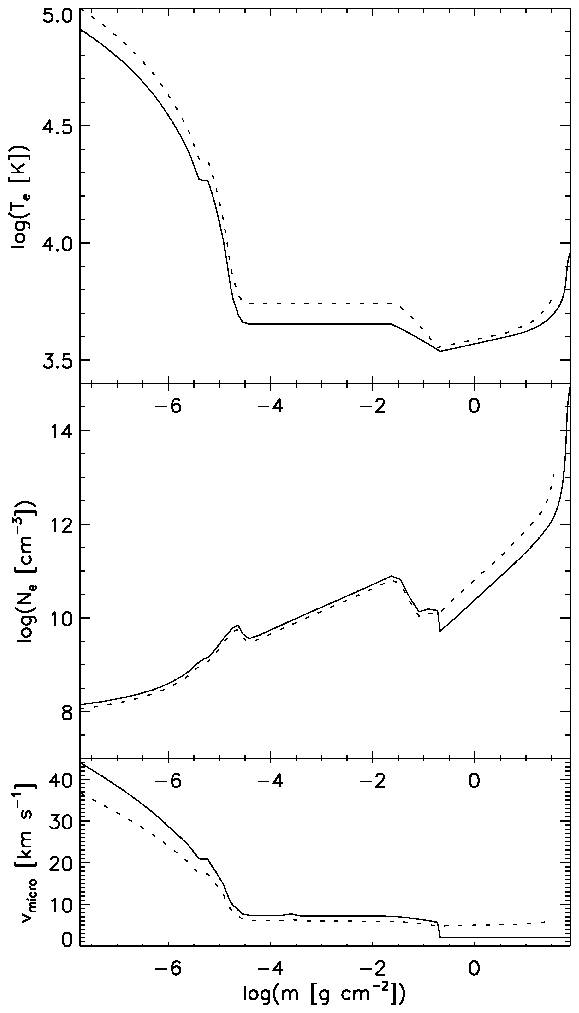}
\caption{Electronic temperature, electronic number density and microturbulent velocity profiles as a function of the integrated mass density $\log m$ of $\beta$~Cet used to reproduce \caII\ triplet line. 
        Dashed line is for model of \citet{Eriksson83}.
        Full line is for a hybrid model between an interpolated MARCS model and a scaled chromosphere of \citet{Eriksson83}'s model.}
\label{Atm}
\end{figure}

The flux and intensity profiles are computed using the non-local thermal equilibrium (NLTE) radiative transfer code MULTI2.2 \citep{Carlsson86} with the \caII\ model atom described in \citet{Merle11} and a model atmosphere corresponding to $\beta$~Cet. We notice that under the assumption of local thermal equilibrium (LTE), the \caII\ triplet lines appear with a core in emission if we use a model atmosphere with a chromosphere. This justifies the use of NLTE radiative transfer in correctly treating the line formation. The MULTI2.2 code provides the contribution functions of the flux showing where the different parts of the line are formed in the atmosphere. We see in Fig.~\ref{CaTcntrb} that the continuum is formed in the deep photosphere and the cores of the \caII\ triplet lines between $\log m = -2$ and $-4$ in the atmosphere, corresponding to the mean chromosphere (see top panel of Fig.~\ref{Atm}). We note that the contribution function in the continuum for the \citet{Eriksson83} model is truncated. This is because their model does not consider sufficient depths of the photosphere, as shown on the top panel of Fig.~\ref{Atm}. The maximum value of the contribution functions in the line cores of the \caII\ triplet corresponds to the mean chromosphere characterized by a constant temperature (see Fig.~\ref{Atm}), i.e. the plateau temperature $T_{\rm p}$. The plateau temperature was a parameter of the model of \citet{Eriksson83} set to fit the \mgII\ k line flux from the IUE observations. From Fig.~\ref{CaTcntrb}, we conclude that the chromosphere is optically thick in the core of the \caII\ triplet lines. Using the interpolated MARCS model without chromosphere, the line core contribution functions are zero, confirming this conclusion. In addition, we note that in Fig.~\ref{CaTcntrb}, the \caII/849~nm line is formed more deeply in the chromosphere than the other two lines. This prediction is clearly supported by our interferometric measurements (see Table \ref{tableextension}). We emphasize that we measure similar relative chromosphere extents in the three components of the \caII\ triplet for the only other star, $\eta$~Cet, for which we have interferometric observations in the core of the \caII\ triplet.

\begin{figure}[]
\includegraphics[width=9cm]{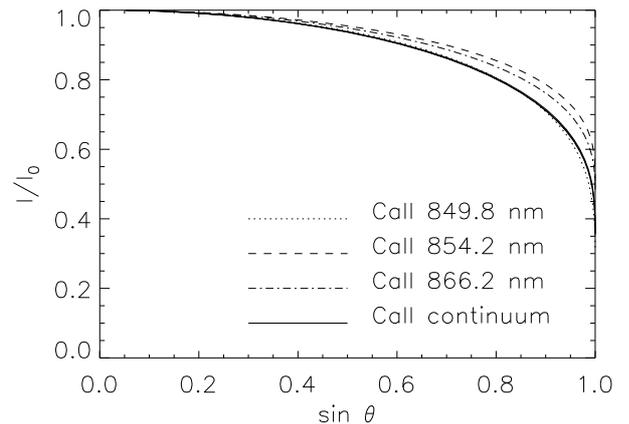}
\caption{NLTE computed limb-darkening laws for \caII\ triplet line cores ($\Delta\lambda = 0.25\ \AA$) and continuum with the hybrid model for $\beta$ Cet as a function of $\sin \theta =\sqrt{1-\mu^2}$.}
\label{LD_CaT}
\end{figure}

\begin{figure*}[h]
\hbox{
\includegraphics[width=6cm]{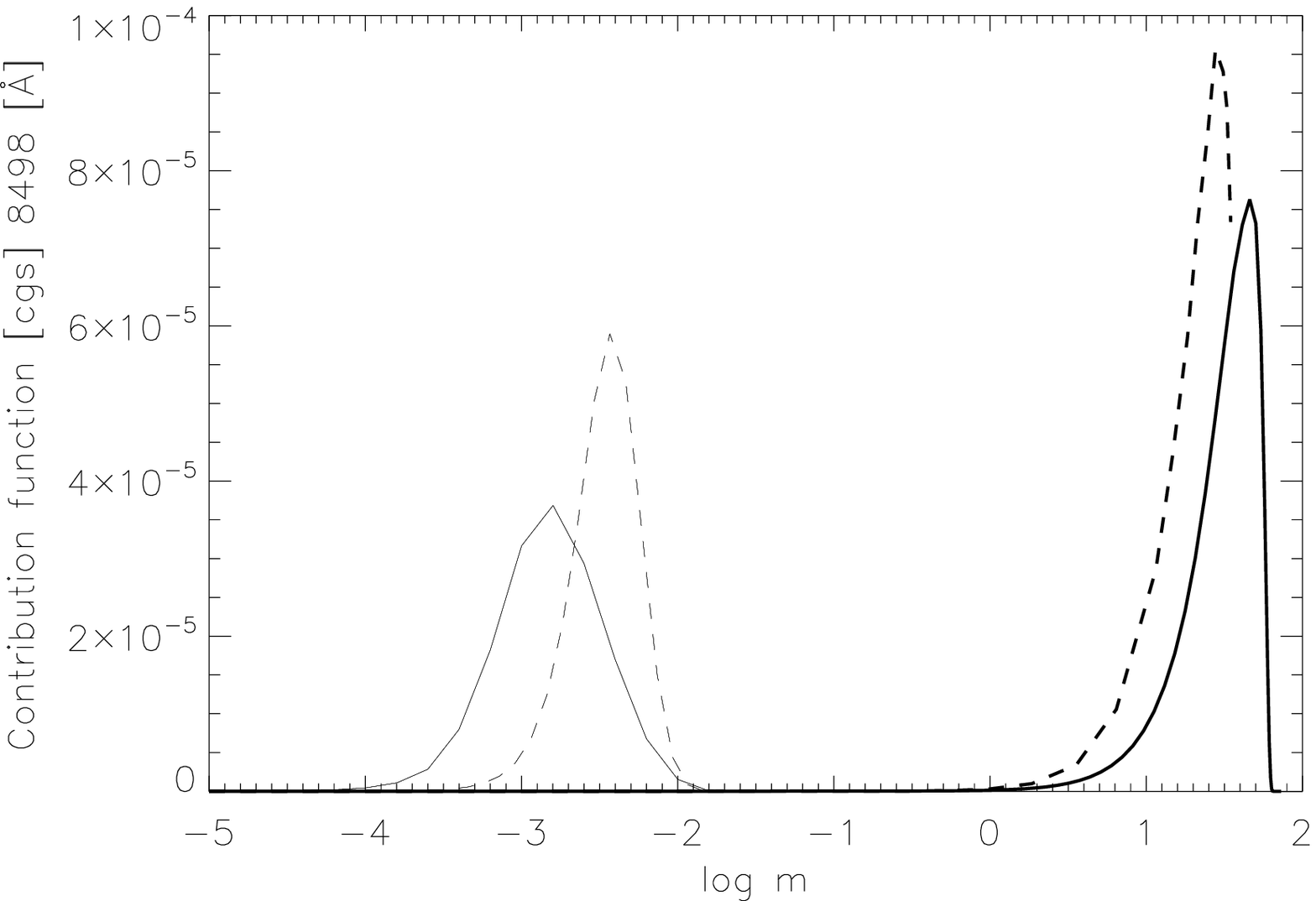}
\includegraphics[width=6cm]{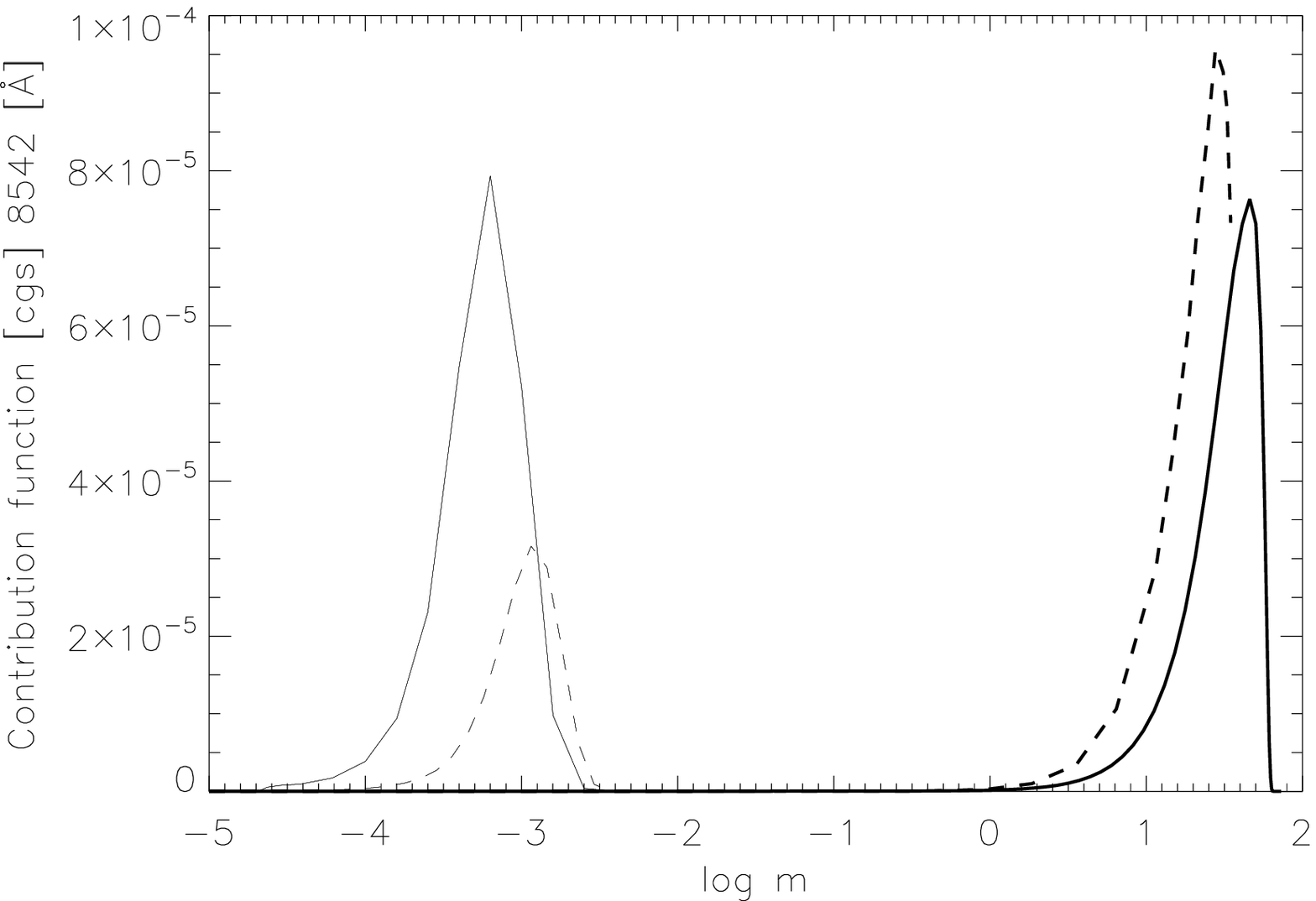}
\includegraphics[width=6cm]{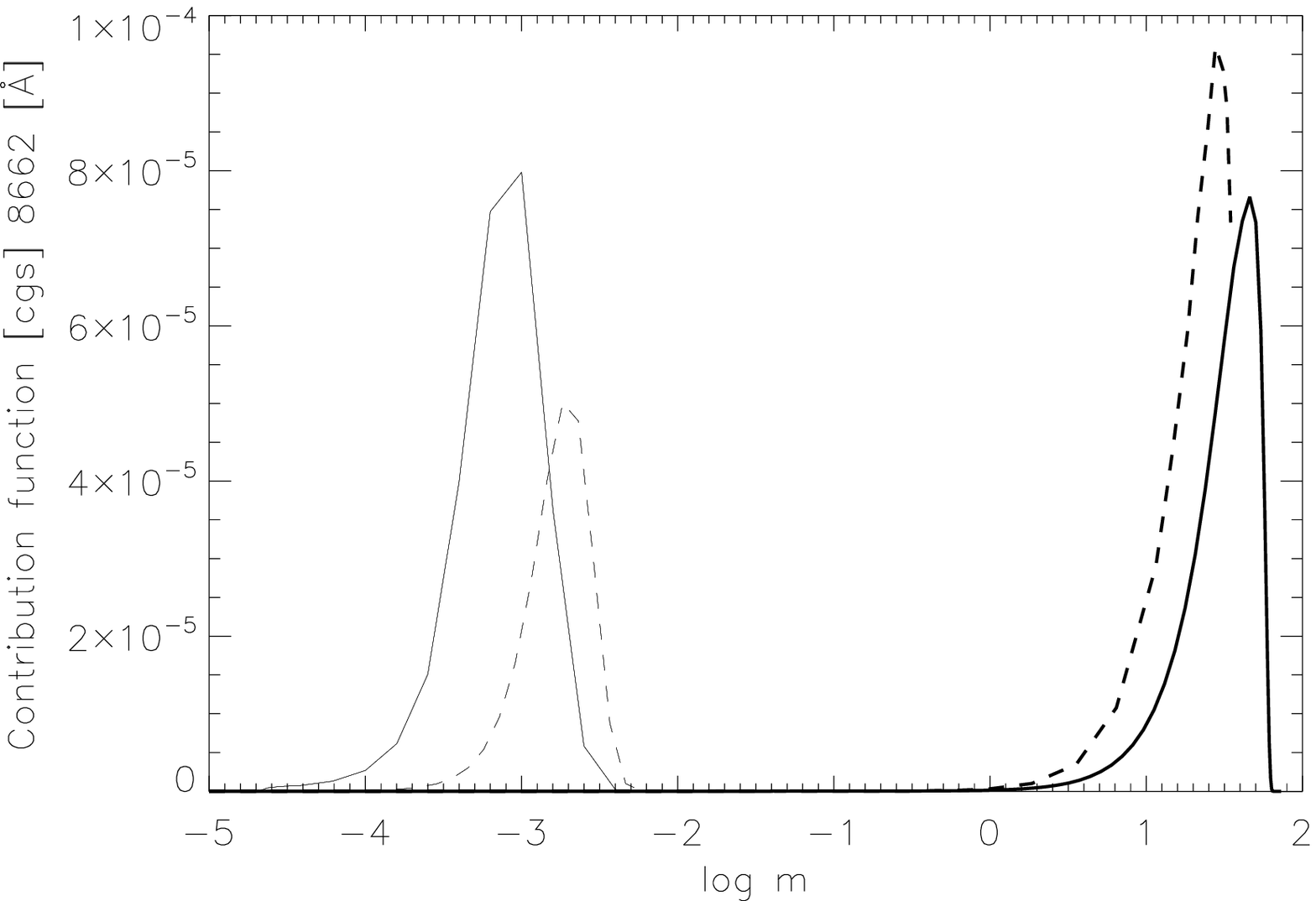}
} 
\caption{Contribution functions to both the core (thin lines) and the continuum (thick lines) of the \caII\ triplet line profiles in $\beta$~Cet. 
        The dashed lines represent the \citet{Eriksson83} model atmosphere.
        The full lines are for the hybrid model.}
\label{CaTcntrb}
\end{figure*}

We use MULTI2.2 to compute the LD laws presented in Fig.~\ref{LD_CaT} for the core of the \caII\ triplet lines and for the close continuum. For this purpose, we modified the line transfer in MULTI2.2 code in order to compute 32 directions ($\mu = \cos \theta$) between 0 and 1. Our result validates the approximation of the UD model for deriving the chromospheric radius, shown in Table~\ref{tableextension}. The computations invalidate the possibility that the line cores could be brighter in the limb than in the center-disk, as we thought at the beginning of the study. The shape of the LD law in the chromospheric line core is similar to the shape of the H$_\alpha$ Balmer lines of the Sun derived by \citet{white62}. 
We then use our model to derive $\theta_{LD}$. For that, we fit to each spectral line a third-order polynomial law in $\mu$ to the intensity profile presented in Fig.~\ref{LD_CaT}. These coefficients are then used to derive $\theta_{LD}$ by fitting a third-order LD law to our visibility measurements. We find that the radius of each \caII\ line core is larger by the same factor of $\sim 5\%$ ($\theta_{LD}(849~{\rm nm})=6.11\pm 0.21~{\rm mas}$, $\theta_{LD}(854~{\rm nm})=6.72\pm 0.24~{\rm mas}$, and $\theta_{LD}(866~{\rm nm})=6.91\pm 0.25~{\rm mas}$) than the $\theta_{UD}$ presented in Table~\ref{tableextension}.

\begin{figure*}[h]
\hbox{
\includegraphics[width=6cm]{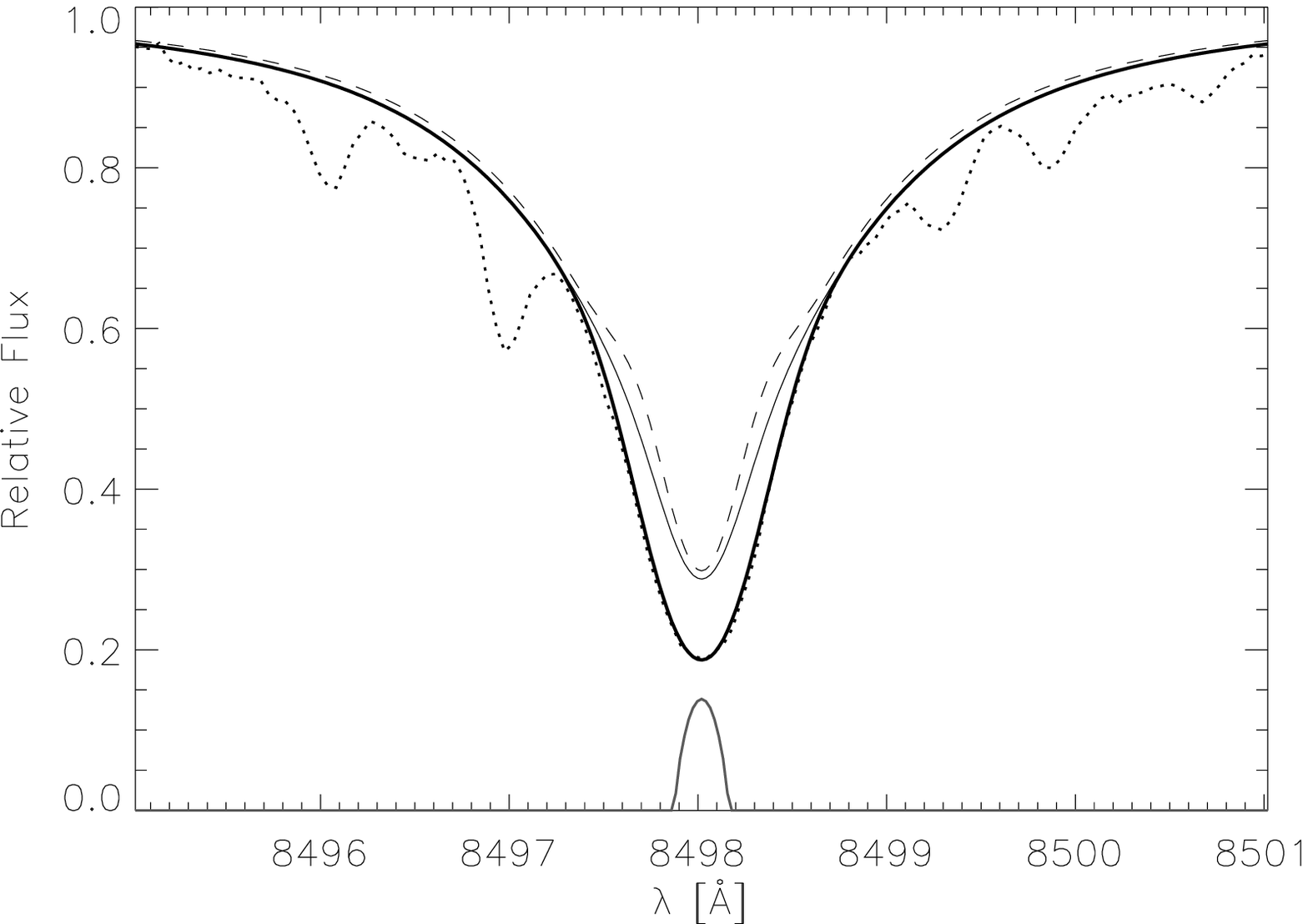}
\includegraphics[width=6cm]{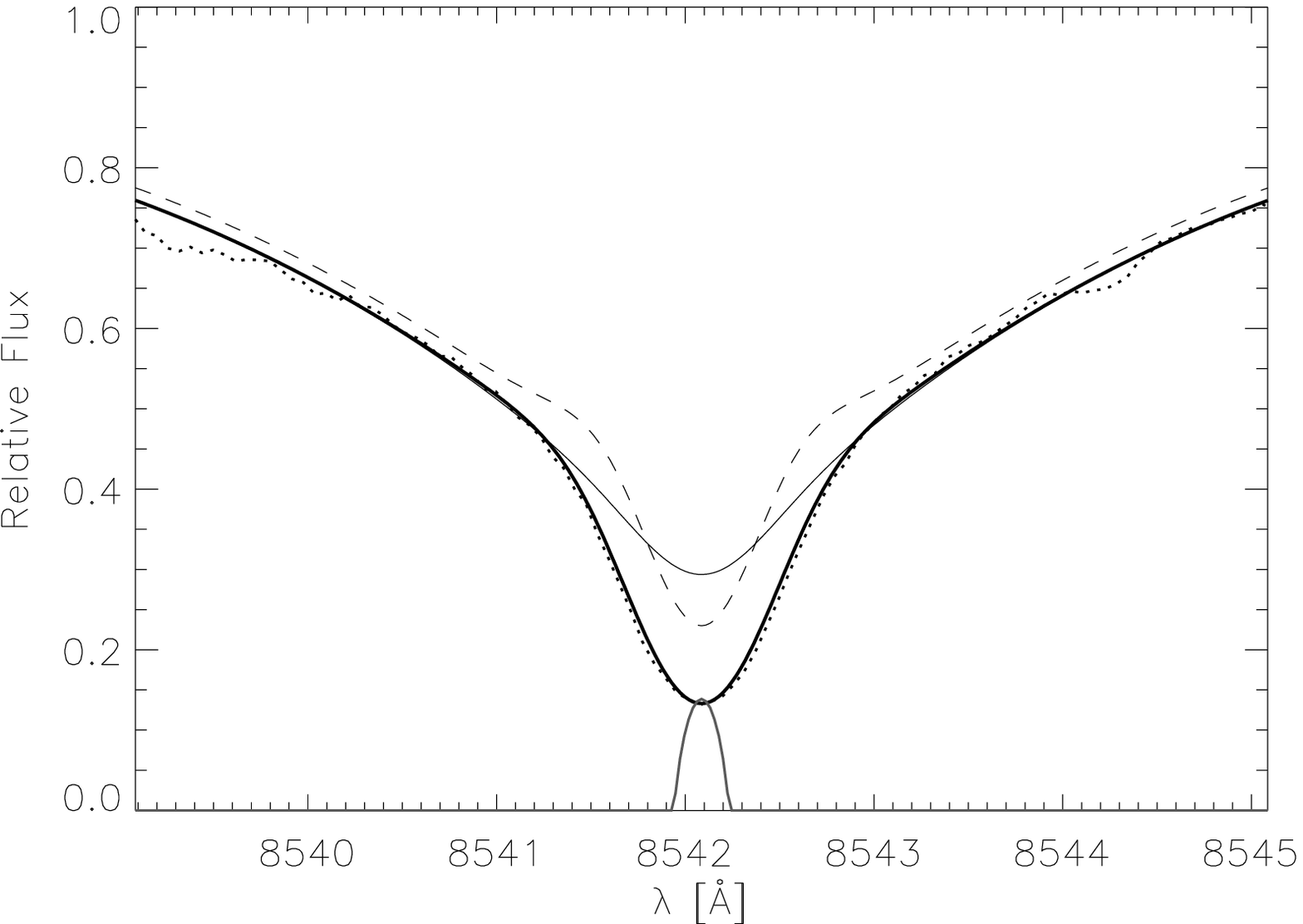}
\includegraphics[width=6cm]{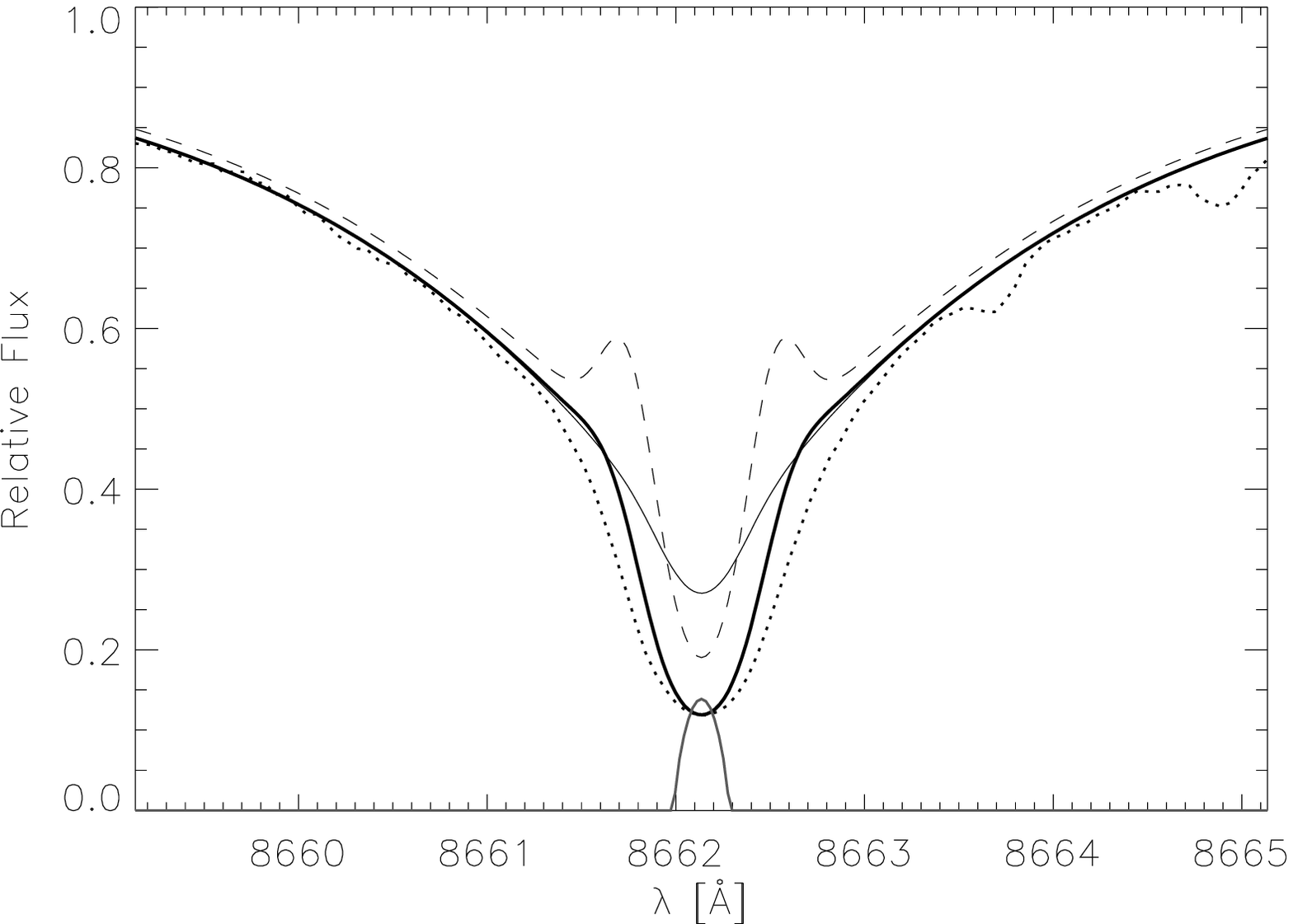}
} 
\caption{Fits of the NLTE \caII\ triplet line profiles in $\beta$~Cet. 
        The dotted line is the observations from NARVAL at TBL with a resolution of $\sim 65,000$.
        The dashed line is the \citet{Eriksson83} model atmosphere with a chromosphere ($T_{{\rm p}}=5,500$~K).
        The thin full line is the interpolated MARCS model.
        The thick full line is our hybrid model with a chromosphere ($T_{{\rm p}}=4,500$).
        The grey lines below the spectral line profiles show the rotation profiles with a projected rotational velocity of 5~km~s$^{-1}$.}
\label{CaTline}
\end{figure*}

Finally, we compute NLTE \caII\ line profiles for three model atmospheres: (1) the model of \citet{Eriksson83} with the incorrect surface gravity, (2) the interpolated MARCS model without chromosphere, and (3) the hybrid model.
The NLTE flux profiles of \caII\ triplet lines are compared with the observations of $\beta$ Cet from the NARVAL instrument at TBL\footnote{available at: http://magics.bagn.obs-mip.fr} ($R \sim 65,000$; $S/N \sim 500$).
The projected rotational velocity of $\beta$~Cet is estimated to be $\sim 5$km~s$^{-1}$ \citep{Massarotti08}.
We included the convolution of the intrinsic flux profile with the rotation profile before degrading to the spectrograph resolution.   
Using the model of \citet{Eriksson83}, we find evidence of emission near the core in Fig.~\ref{CaTline}, especially for the {\caII}/866~nm line.  
We interpret this emission as an indication of a higher temperature $T_{{\rm p}}$. As an exercice, we scale the chromosphere and transition part (as proposed by \citealt{Eriksson83}) in temperature until the emission parts are removed and the line cores are fitted. 
This leads to a  plateau temperature of $T_{{\rm p}} = 4,500$~K, instead the value of $T_{{\rm p}} = 5,500$~K assumed in the original model.
To maintain the physical conditions of the chromosphere, we also scaled the electronic number density in order to keep the electronic pressure constant.
Moreover, we also increased the microturbulent velocity profile of \citet{Eriksson83}'s chromosphere by 20~\% in order to enlarge the gaussian core of the lines. 
The hybrid model with $T_{{\rm p}} = 4,500$~K and larger microturbulent velocity is compared with the model of \citet{Eriksson83} in Fig.~\ref{Atm}.
As shown in Fig. \ref{CaTline}, we satisfactorily reproduce the line cores, but only when applying a macroturbulent parameter of 4 and 11 for the {\caII}/849 nm and {\caII}/854 nm lines, respectively to the three models. However, we note that our model does not perfectly reproduce the line core for the {\caII}/866 nm line.
For strong lines such as the \caII/ triplet, the weak projected rotational velocity of $\beta$~Cet has no effect on the line cores.
We note the difference in the line core between an interpolated MARCS model without a chromosphere and the hybrid model with a chromosphere. 
This difference can be used as an indicator of a chromosphere and stellar activity (e.g. \citealt{Andretta05} and \citealt{Busa07}).

\section{Conclusion}
By analyzing interferometric measurements derived from CHARA/VEGA data, we have determined the physical extents of the chromosphere of late-type giant stars. For the first time, we have measured the position of the area in the atmosphere where \caII\ triplet and H$_{\alpha}$ lines formed and found that these chromospheres are relatively extended, between 16\% and 47\% of the stellar radius of the corresponding photospheres. Surprisingly, we have found that four of the program stars exhibit anomalous phases of the interferometric fringes that can be interpreted as an inhomogeneity in the chromospheric part of the atmosphere. More observations with a more complete $(u,v)$ coverage are needed to explain the detected asymmetry of these chromospheres. 

In addition to this chromospheric study, we have derived the fundamental parameters of seven cool giant stars. From the stellar radius measured in the continuum and the corresponding bolometric flux, we have deduced the effective temperatures of the program stars, which are in excellent agreement with T$_{\rm eff}$ scale laws found in the literature for such K giant stars. 

We have also developed a semi-empirical model atmosphere of $\beta$ Cet. This model has helped us to define the limb-darkening law to apply to the core of the \caII\ triplet lines. We have found that it is similar to the one found for the Sun. We have used this model to define where the cores of the \caII\ triplet lines are formed by analyzing the contribution functions. We have been able to confirm that these lines are formed at the mean chromosphere temperature and that the {\caII}/849~nm line is formed significantly deeper within the atmosphere than the two other \caII\ triplet lines. This result is in very good agreement with the interferometric measurements for $\beta$ and $\eta$ Cet. Finally, we have computed synthetic \caII\ line profiles, which have been compared to an observed high resolution spectrum of $\beta$ Cet. We have found that the temperature plateau (as proposed by \citealt{Eriksson83}) of the chromosphere model is too high to reproduce the line cores. 

On the basis of these first interferometric results on the spatial structure of the chromosphere of K giant stars, it now seems possible to construct reliable chromospheric models by combining spectroscopic and interferometric observations. 

\begin{acknowledgements}
VEGA is a collaboration between CHARA and
OCA / LAOG / CRAL / LESIA that has been supported by the French 
programs PNPS and ASHRA, by INSU and by the R\'{e}gion
PACA. The CHARA Array is operated with support from the
National Science Foundation through grant AST-0908253, the W. M.
Keck Foundation, the NASA Exoplanet Science Institute, and from
Georgia State University. This research has made use of the Jean-
Marie Mariotti Center LITpro and SearchCal services co-developed
by CRAL, LAOG and FIZEAU, and of CDS Astronomical Databases
SIMBAD and VIZIER. Finally, we thank the anonymous referee for providing 
pertinent advices that helped improving this paper.
\end{acknowledgements}

\bibliographystyle{aa}
\bibliography{biblio}

\end{document}